\documentclass[twocolumn,showpacs,preprintnumbers,amsmath,amssymb,prb,superscriptaddress]{revtex4-1}
\usepackage{epsfig}
\usepackage{graphicx}
\usepackage{dcolumn}
\usepackage{bm}
\usepackage{color} 

\begin{document}

\title{
Optical signature of Weyl electronic structures in tantalum pnictides \\Ta$Pn$ ($Pn=$~P, As)
}
\author{Shin-ichi Kimura}
\email{kimura@fbs.osaka-u.ac.jp}
\affiliation{Graduate School of Frontier Biosciences, Osaka University, Suita 565-0871, Japan}
\affiliation{Department of Physics, Graduate School of Science, Osaka University, Toyonaka 560-0043, Japan}
\author{Hiroko Yokoyama}
\affiliation{Department of Physics, Graduate School of Science, Osaka University, Toyonaka 560-0043, Japan}
\author{Hiroshi Watanabe}
\affiliation{Graduate School of Frontier Biosciences, Osaka University, Suita 565-0871, Japan}
\affiliation{Department of Physics, Graduate School of Science, Osaka University, Toyonaka 560-0043, Japan}
\author{J\"org Sichelschmidt}
\affiliation{Max Planck Institute for Chemical Physics of Solids, 01187 Dresden, Germany}
\affiliation{Graduate School of Frontier Biosciences, Osaka University, Suita 565-0871, Japan}
\author{Vicky S\"u\ss}
\affiliation{Max Planck Institute for Chemical Physics of Solids, 01187 Dresden, Germany}
\author{Marcus Schmidt}
\affiliation{Max Planck Institute for Chemical Physics of Solids, 01187 Dresden, Germany}
\author{Claudia Felser}
\affiliation{Max Planck Institute for Chemical Physics of Solids, 01187 Dresden, Germany}
\date{\today}
\begin{abstract}
To investigate the electronic structure of Weyl semimetals Ta$Pn$ ($Pn=$~P, As), optical conductivity [$\sigma(\omega)$] spectra are measured over a wide range of photon energies and temperatures, and these measured values are compared with band calculations.
Two significant structures can be observed: a bending structure at $\hbar\omega\sim$~85~meV in TaAs, and peaks at $\hbar\omega\sim$~50~meV (TaP) and $\sim$~30~meV (TaAs).
The bending structure can be explained by the interband transition between saddle points connecting a set of $W_2$ Weyl points.
The temperature dependence of the peak intensity can be fitted by assuming 
the interband transition between saddle points connecting a set of $W_1$ Weyl points.
Owing to the different temperature dependence of the Drude weight in both materials, it is found that the Weyl points of TaAs are located near the Fermi level, whereas those of TaP are further away.
\end{abstract}

%
%
%
%
\maketitle
%
\section{Introduction}
%
Recently, some materials such as graphene and the edge states of topological insulators have been observed to exhibit linear band dispersions that can be explained by the Dirac equation. As these materials have high mobility owing to their massless carriers and high Fermi velocities, 
they are expected to be suitable for applications involving high-speed devices~\cite{Hills2015}.
Graphene has a space inversion symmetry and degenerate bands with up- and down-spins~\cite{Neto2009}.
However, in some materials with strong spin-orbit coupling, the up- and down-spin bands split when the space inversion symmetry is broken; one example of such a material is BiTeI~\cite{Ishizaka2011}.
Weyl semimetals (WSMs) have linear band dispersions and large spin-orbit coupling~\cite{Bernevig2015}.
The crossing points of these linear band dispersions, namely Weyl points, are not located at highly symmetric points, but at a mirror position relative to the symmetry plane.
Fermi arcs have been observed on the surfaces of WSMs; these occur because the Weyl points have a Berry phase and non-trivial topological numbers~\cite{Xu2015-1,Xu2015-2}.
WSMs exhibit a large positive magneto-resistance in the $I \perp B$ configuration, but a large negative magneto-resistance when $I \parallel B$, 
namely an Adler--Bell--Jackiw chiral anomaly~\cite{Du2016,Zhang2015,Huang2016}.

Tantalum pnictides, TaP and TaAs, are WSMs~\cite{Weng2015}
in which the crystal has a body-centered-tetragonal NbAs-type structure with a nonsymmorphic space group of $I4_1md$ (No.~109).
According to band calculations, there are 12 pairs of Weyl points (24 points in total)~\cite{Weng2015,Lee2015}.
Four of these pairs are located near the $\Gamma-\Sigma$ line (namely $W_1$) and the other eight pairs are located near the $\Gamma-N$ line (namely $W_2$).
These two kinds of Weyl points have different characteristics, with the $W_2$ points of TaP (TaAs) expected to be located far from (just at) the Fermi level $E_F$ suggested by band calculations~\cite{Lee2015}.
The bulk electronic structures of these materials have been investigated by angle-resolved photoelectron spectroscopy using soft-X-ray excitation light~\cite{Lv2015,NXu2016}.
These experiments revealed linear band dispersions near the Weyl points, but the energy position of the Weyl points has not yet been clarified because of the limited energy resolution.

The bulk electronic structure near $E_F$ can be observed by optical conductivity [$\sigma(\omega)$] spectra in the infrared region.
To date, the $\sigma(\omega)$ spectra of three-dimensional (3D) Dirac materials such as Cd$_3$As$_2$~\cite{Neubauer2016}, Nd$_2$(Ir$_{0.98}$Rh$_{0.02}$)$_2$O$_7$~\cite{Ueda2012}, Eu$_2$Ir$_2$O$_7$~\cite{Sushkov2015}, ZrTe$_5$~\cite{Chen2015}, quasicrystals~\cite{Timusk2013}, YbMnBi$_2$~\cite{Chinotti2016,Chaudhuri2017}, and TaAs~\cite{BXu2016} have been investigated.
In all of these materials, $\sigma(\omega)$ is proportional to the photon energy ($\omega$-linear component), which suggests interband transitions in the 3D Dirac bands~\cite{Hosur2012}.
Ideally, the Drude weight should disappear at the temperature of 0~K if the Dirac point is located at $E_F$.
However, in real materials, the Dirac points are slightly away from $E_F$, and the Drude weight remains at the lowest temperatures.

According to theoretical calculations, when the Weyl point is located far from $E_F$, 
not only does the Drude weight remain at zero temperature, but also a step appears in the $\omega$-linear component of the $\sigma(\omega)$ spectra~\cite{Ashby+Carbotte2014,Tabert+Carbotte+Nicol2016}.
The energy of the step is twice that of the Weyl point to $E_F$.
This step-feature might be obscured by another spectral feature that is expected in the presence of a pair of Weyl-point cones connected by saddle points forming a two-leg-like band structure~\cite{Tabert+Carbotte2016}.
If states near these saddle points are involved in the interband transitions, the slope of the $\omega$-linear component of the $\sigma(\omega)$ spectra changes, i.e., a ``bending point'' occurs.
These $\sigma(\omega)$ structures, ``step'' or ``bending'' are characteristic features to identify the electronic structure of WSMs.

In this paper, we have measured the temperature dependence of the $\sigma(\omega)$ spectra for the typical WSMs TaP and TaAs, and compared the $\sigma(\omega)$ measurements with the corresponding band calculations to investigate the electronic structure near $E_F$.
The peaks of obtained $\sigma(\omega)$ spectra in the photon energy range above 300~meV can be 
roughly explained by electronic band structure calculations without the effects of renormalization and electron correlation.
From the spectral shape below 200~meV, we obtained detailed information about the band structure near $E_F$, the temperature dependence of the carrier mass, and the energy positions of the Weyl points and saddle points connecting the two paired cones of Weyl points.
A peak structure that cannot be explained by the calculated $\sigma(\omega)$ spectra appears at about 50~meV in TaP and at 30~meV in TaAs.
This structure may originate from an electron-hole pair excitation between the saddle points with a temperature dependence arising from a thermal screening (Pauli blocking) effect.

\section{Experimental and calculation methods}
%
\begin{figure}[t]
\begin{center}
\includegraphics[width=0.45\textwidth]{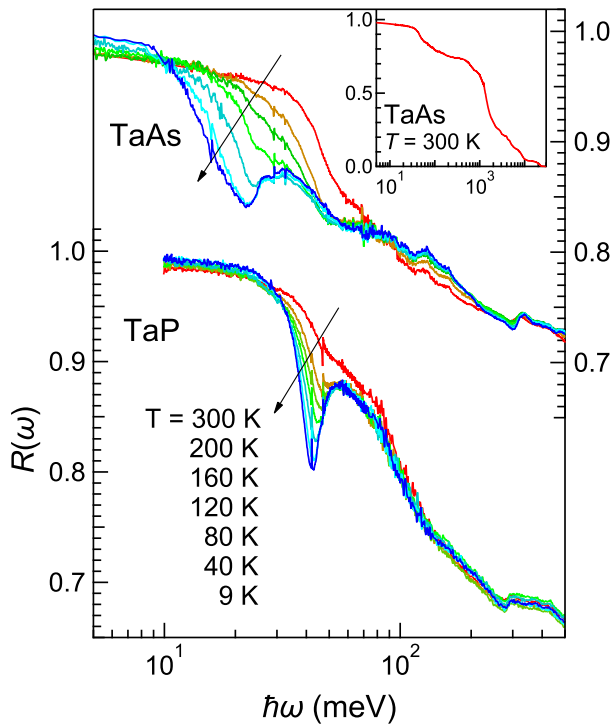}
\end{center}
\caption{
(Color online)
Temperature-dependent reflectivity [$R(\omega)$] spectra of TaP and TaAs in the energy region of 5--500~meV.
The inset shows the $R(\omega)$ spectrum of TaAs at 300~K in the whole measured region up to 30~eV.
}
\label{fig:reflectivity}
\end{figure}

Single-crystalline samples of Ta$Pn$ with a size of approximately 1~mm in diameter were synthesized by the chemical vapor growth method.
We measured the optical reflectivity [$R(\omega)$] spectra of both the as-grown and polished surfaces of Ta$Pn$.
However, the polished surfaces have normal metallic-like spectra with no temperature dependence, 
whereas the as-grown surfaces exhibit clear spectral features with a large temperature dependence consistent with those of other physical properties.
Therefore, the as-grown surfaces appear to have bulk properties, 
and, thus, only the spectra for the as-grown surfaces are used hereafter to discuss the electronic structure.
Near-normal incident $R(\omega)$ spectra were acquired over a wide photon-energy $\hbar\omega$ range of $8~{\rm meV}-30~{\rm eV}$ to ensure accurate Kramers--Kronig analysis (KKA)~\cite{Kimura2013}.
Michelson- and Martin--Puplett-type rapid-scan Fourier spectrometers were used at $\hbar\omega$ regions of $8~{\rm meV}-1.5~{\rm eV}$ and $3-20~{\rm meV}$, respectively, with a feedback positioning system to maintain the overall uncertainty level within $\pm$0.3\% over a temperature range of $T=9-300~{\rm K}$~\cite{Kimura2008}.
To obtain the absolute $R(\omega)$ values, an {\it in-situ} evaporation method was adopted.
The obtained temperature-dependent $R(\omega)$ spectra of TaP and TaAs are plotted in Fig.~\ref{fig:reflectivity}.
The $R(\omega)$ spectrum for 1.2--30~eV was measured at 300~K using synchrotron radiation, and this was connected to the spectra for $\hbar\omega \leq 1.5$~eV for the KKA.
All measurements were performed along the $ab$-plane, i.e., the electric vector is proportional to the $a$ and $b$ axes.
To obtain $\sigma(\omega)$ via $R(\omega)$ KKA, the spectra were extrapolated below 3~meV with a Hagen--Rubens function and above 30~eV with a free-electron approximation~\cite{DG}, $R(\omega) \propto \omega^{-4}$.

The obtained $\sigma(\omega)$ spectra were compared with those given by local-density approximation band structure calculations including spin-orbit coupling using the {\sc Wien2k} code~\cite{Wien2k}
(with lattice parameters of $a=b=3.3184~{\rm \AA}$ and $c=11.3388~{\rm \AA}$ for TaP and $a=b=3.4348~{\rm \AA}$ and $c=11.6410~{\rm \AA}$ for TaAs).

\section{Results and Discussion}
\subsection{Electronic structure in the wide energy range up to 1.2~eV}
\begin{figure}[t]
\begin{center}
\includegraphics[width=0.45\textwidth]{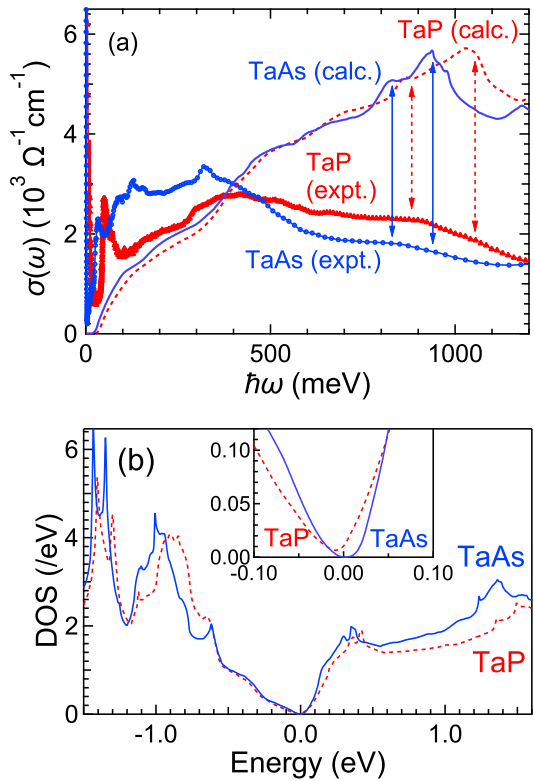}
\end{center}
\caption{
(Color online)
(a) Experimentally obtained optical conductivity [$\sigma(\omega)$] spectra of TaP and TaAs at the temperature of 9~K compared with the spectra obtained from band calculations.
The arrows indicate the corresponding peaks between experimental and calculated $\sigma(\omega)$ spectra.
(b) Calculated density of states (DOS) of TaP and TaAs.
The zero energy corresponds to the Fermi energy ($E_F$).
The inset shows an enlargement near $E_F$.
}
\label{fig:OCwide}
\end{figure}
Figure~\ref{fig:OCwide}(a) shows the experimental $\sigma(\omega)$ spectra of TaP and TaAs at $T$ = 9~K for up to 1.2~eV, together with the theoretical spectra from band calculations.
The whole experimental spectrum of TaP, especially the wide peak at $\hbar\omega\sim$~400~meV and two shoulder structures at $\sim$1000~meV, is slightly shifted to the high-energy side compared with that of TaAs.
These peaks can be reproduced by band calculations, i.e., the 400-meV broad peak corresponds to a broad shoulder at $400-500~{\rm meV}$ and the 1000-meV shoulders can be regarded as the peaks indicated by arrows.
This result suggests that the experimental $\sigma(\omega)$ spectra and the electronic structure can be explained by the band calculations without a renormalization effect, which is usually introduced in strongly correlated electron systems~\cite{Kimura2009}.
It should be noted that the calculated spectral intensity above 500~meV is larger and also the calculated peaks at around 1000~meV are narrower than the experimental ones due to the smaller lifetime broadening in the calculation (1.36~meV).
In addition, the $\sigma(\omega)$ spectrum of TaP in $\hbar\omega\leq$~500~meV has a lower intensity than that of TaAs, which is consistent with the calculated spectra.
The origin of this phenomenon is that the unoccupied density of states (DOS) of TaP is lower than that of TaAs, as shown in Fig.~\ref{fig:OCwide}(b).
The DOS peaks of TaP at around 0.4~eV and 1.5~eV are shifted to the high-energy side by about 0.1~eV compared with those of TaAs.
These differences in the electronic structure between TaP and TaAs are reflected in the $\sigma(\omega)$ spectra.

\subsection{Electronic structure near $E_F$}
\begin{figure}[b]
\begin{center}
\includegraphics[width=0.45\textwidth]{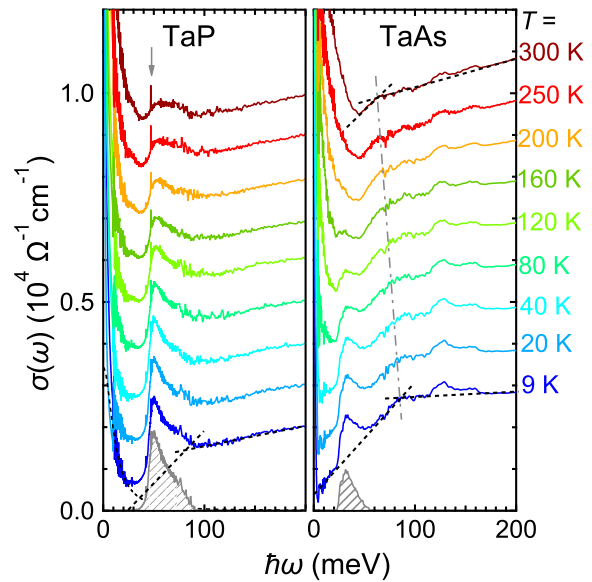}
\end{center}
\caption{
(Color online)
Temperature-dependent optical conductivity [$\sigma(\omega)$] spectra of TaP and TaAs in the photon energy $\hbar\omega$ range below 200~meV.
The baseline of each spectrum is shifted by $10^3~\Omega^{-1}{\rm cm}^{-1}$.
For TaAs, the dashed lines denote two $\omega$-linear components at $T=9~{\rm K}$ and 300~K, and the dot-dashed line is a guide for the temperature dependence of the bending point.
For TaP, the dashed lines are fitted Drude functions and two $\omega$-linear components at $T=~9~{\rm K}$.
In the figure, the energy of the bending point is assumed to be the same as that of TaAs.
Hatched peaks show the remaining $\sigma(\omega)$ spectra given by subtracting the dashed lines from the $\sigma(\omega)$ spectra at 9~K.
The sharp peak of TaP (indicated by an arrow) is caused by an optical phonon; this is not discussed in the present paper~\cite{Xu2017}.
}
\label{fig:OCnarrow}
\end{figure}
We now discuss the electronic structure near $E_F$.
Figure~\ref{fig:OCnarrow} shows the temperature-dependent $\sigma(\omega)$ spectra for $\hbar\omega\leq~200~{\rm meV}$.
The spectral weights of both materials are conserved in the energy range below 200~meV.
TaAs has four significant structures---a Drude peak for $T\geq80~{\rm K}$, a peak at $\sim30~{\rm meV}$ for $T\leq160~{\rm K}$, 
and two $\omega$-linear components below and above 85~meV at 9~K.
With increasing temperature, the bending point at 85~meV (9~K) shifts slightly to the low-energy side, to 60~meV (300~K).
The bending point corresponds to a van Hove singularity of the optical transition between saddle points connecting two Weyl points, as has been previously reported~\cite{Tabert+Carbotte2016}.
TaP has three significant structures---a Drude peak owing to carriers below 30~meV, a peak at about 50~meV, and an $\omega$-linear structure above 100~meV.
These three structures are observed in both materials.
However, the low-energy $\omega$-linear component that clearly appears in TaAs is not observed in TaP.
In TaP, the large 50-meV peak dominates eventually existent spectral features typical for WSMs, i.e., a low-energy $\omega$-linear component or a step in the high-energy $\omega$-linear component~\cite{Ashby+Carbotte2014}.
In the following analysis, we assume the presence of a low-energy $\omega$-linear component below 85~meV, as indicated in the figure.
However, this assumption leads to some ambiguity in evaluating the 50-meV peak intensity (see below). 

\begin{figure}[b]
\begin{center}
\includegraphics[width=0.45\textwidth]{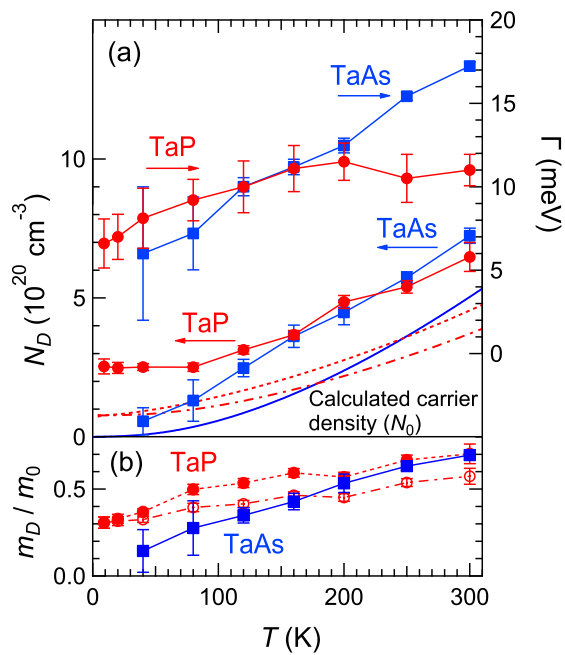}
\end{center}
\caption{
(Color online)
(a) Temperature-dependent Drude parameters of effective carrier density (Drude weight, $N_D$) and damping factor (inverse relaxation time, $\Gamma$) obtained by the fitting of the $\sigma(\omega)$ spectra of Fig.~\ref{fig:OCnarrow}.
The solid, dashed, and dot-dashed lines are the calculated carrier densities ($N_0$) of TaAs, TaP with $E_F=0~{\rm eV}$, and TaP with $E_F=-20~{\rm meV}$, respectively, using the calculated DOS in Fig.~\ref{fig:OCwide}(b).
(b) Temperature-dependent effective mass ($m_D/m_0=N_0/N_D$) of TaAs (solid square--solid line), TaP with $E_F=0~{\rm meV}$ (solid circle--dashed line), and TaP with $E_F=-20~{\rm meV}$ (open circle--dot-dashed line), derived from the comparison of experimental $N_D$ with calculated $N_0$ in (a).
}
\label{fig:Drude}
\end{figure}
In TaP, the Drude peak is observed at all temperatures, whereas for TaAs, the Drude peak is suppressed to zero energy for $T\leq80~{\rm K}$.
This difference can also be observed in the $R(\omega)$ spectra (Fig.~\ref{fig:reflectivity}), i.e., 
in contrast to the weak temperature dependence of the energy of the local minimum (plasma edge) at about 40~meV in TaP, the plasma edge at 50~meV at 300~K rapidly shifts to the low-energy side (20~meV at 9~K) with decreasing temperature.
To clarify the difference in these materials' temperature dependence quantitatively, the Drude peaks were fitted using the following Drude function:
\begin{equation}
\sigma(\omega)=\frac{N_{D}e^2}{m_0} \frac{\Gamma}{(\hbar\omega)^2+\Gamma^2}.
\label{eq:Drude}
\end{equation}
Here, $e$ and $m_0$ are the electron charge and rest mass, respectively.
The obtained $N_D$ and $\Gamma$ are the fitting parameters for the Drude weight and damping factor, respectively. These are plotted as a function of temperature in Fig.~\ref{fig:Drude}(a).

For TaAs, both $N_D$ and $\Gamma$ decrease almost linearly with decreasing temperature.
When the Weyl points are located at $E_F$, the Drude weight is expected to behave in this way because of the suppression of thermal broadening.
As the electrons near the Weyl points contribute to the Drude weight, the carrier mass then becomes small.
Since mobility is expressed as $e/(m_{eff}\Gamma)$, the mobility becomes large with decreasing temperature owing to both of the decreasing effective carrier mass $m_{eff}$ and the decreasing $\Gamma$.
Note that the Weyl points of TaAs deviate slightly from $E_F$, because the Drude component still remains at the lowest temperature.

Our data for TaAs are similar to those in previous reports~\cite{BXu2016}, although the following differences were observed:
First, the previous data have a Drude peak below 100~cm$^{-1}$ (12~meV) at $T=5~{\rm K}$, but the Drude component in our spectrum at the lowest temperature of 9~K is strongly suppressed below 3~meV (24~cm$^{-1}$).
This suggests that the sample used to obtain the previous data~\cite{BXu2016} had a higher carrier density than ours at low temperatures.
Second, a clear peak appears at about 30~meV (240~cm$^{-1}$) in our data, in contrast to a broad shoulder in the previous data.

For TaP, $N_D$ decreases linearly with decreasing temperature from 300~K to 120~K, consistent with the behavior of TaAs, but remains constant below 80~K.
This implies that the Weyl points of TaP are located far from $E_F$ and/or non-Weyl bands are located on $E_F$.
According to the inset of Fig.~\ref{fig:OCwide}(b), the DOS at $E_F$ of TaP is finite, in contrast to the almost-zero DOS of TaAs.
This remaining DOS of TaP mainly originates from non-Weyl bands.
Therefore, the difference in the temperature dependence of the $\sigma(\omega)$ spectra of TaP and TaAs originates from their different band characteristics near $E_F$.
This conclusion is also supported by the fact that $\Gamma$ in TaP is almost constant above 160~K, in contrast to the linear temperature dependence of TaAs.

In Figure~\ref{fig:Drude}(a), the slope of $N_D$ is sample-dependent, that is, $N_D$ is higher in TaP than in TaAs below 200~K, but the order is reversed above 250~K.
The sample-dependence of the temperature change in $N_D$ can be explained by a thermal broadening of DOS near $E_F$, namely, the DOS at $E_F$ of TaP is higher than that of TaAs, 
but in the energy regions above $50~{\rm meV}$ and below $-50~{\rm meV}$ from $E_F$, the DOS of TaAs is higher than that of TaP.

To confirm these considerations, the temperature dependence of the carrier density [$N_0(T)$] was calculated using the DOS [$D(E)$], 
as shown in Fig.~\ref{fig:OCwide}(b), by means of the following formula:
\begin{equation}
N_0(T)=\int{D(E)\left(-\frac{\partial f_{FD}(E,T)}{\partial E}\right)dE},
\label{eq:N_0}
\end{equation}
where $f_{FD}(E,T)=1/[\exp\frac{E-\mu}{k_BT}+1]$ is the Fermi--Dirac distribution function, where $\mu$ is the chemical potential.
The obtained $N_0(T)$s for TaAs, TaP with $E_F=0~{\rm eV}$, and TaP with $E_F=-20~{\rm meV}$ are plotted in Fig.~\ref{fig:Drude}(a).
The calculation for TaP with $E_F=-20~{\rm meV}$ is related to the analysis of the Weyl points, and is discussed later.
The temperature dependence of $N_0(T)$ in both materials is roughly consistent with $N_D$, i.e., at low temperatures, $N_0(T)$ becomes finite in TaP but almost zero in TaAs.
However, the value and slope of $N_0$ are not the same as the experimental $N_D$.
The difference between $N_0$ and $N_D$ is considered to originate from the effective carrier mass ($m_D$) with the relation $m_D/m_0=N_0/N_D$.
From this formula, the temperature dependence of the effective mass $m_D/m_0$ can be evaluated as shown in Fig.~\ref{fig:Drude}(b).
For TaAs, $m_D/m_0$ of about 0.8 at 300~K monotonically decreases with decreasing temperature to 0.15 at 9~K.
For TaP, the value of $m_D/m_0$ at 300~K is similar to that of TaAs, but gradually decreases down to 80~K, below which it rapidly decreases to 0.3 at 9~K.
The mass reduction ratio of TaAs is half that of TaP, suggesting that the TaAs carriers mainly originate from Weyl bands, whereas those of TaP originate from both the Weyl bands and the normal conduction bands.

\begin{figure}[b]
\begin{center}
\includegraphics[width=0.45\textwidth]{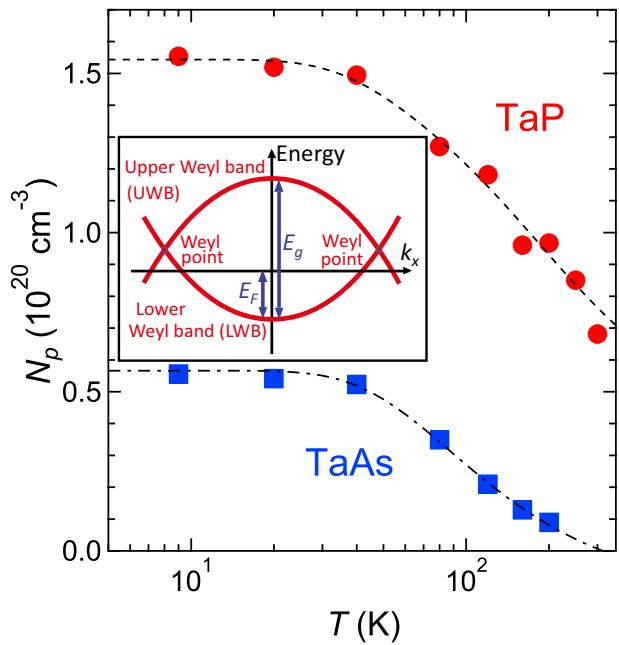}
\end{center}
\caption{
Temperature dependence of peak intensities [effective electron number ($N_{p}$)] of TaP and TaAs at around 50 and 30~meV, respectively.
The error of $N_p$ is less than the symbol size.
The dashed and dot-dashed lines are fitting curves assuming the temperature dependence of interband transition expressed by Eq.~\ref{eq:Pauliblocking}.
The obtained parameters are listed in Table~\ref{table:fitting}.
(Inset) Schematic electronic structure near a set of Weyl points.
There are two bands connecting a set of Weyl points, namely, the upper Weyl band (UWB) and the lower Weyl band (LWB).
$E_F$ and $E_g$ are the Fermi energy and the energy difference between LWB and UWB, respectively.
}
\label{fig:Peak}
\end{figure}
\begin{table}[t]
\caption{
Parameters obtained from a nonlinear least-squares fitting method of the temperature-dependent $N_p$ in Fig.~\ref{fig:Peak} by Eq.~\ref{eq:Pauliblocking}.
$N_{p0}$, $E_F$, $E_g$, and $c$ indicate the peak intensity at zero temperature, the Fermi energy, the energy difference between LWB and UWB, and constant background, respectively.
}
\begin{tabular}{@{\hspace{\tabcolsep}\extracolsep{\fill}}ccc} \hline
				& TaP		& TaAs		\\ \hline
$N_{p0}$ [$10^{19}~{\rm cm}^{-3}$]	& $16.9\pm0.5$	& $7.16\pm0.01$	\\
$E_F$ [meV]			& $56.8\pm5.0$	& $11.6\pm0.3$	\\
$E_g$ [meV]			& $69.1\pm5.4$	& $23.3\pm0.1$	\\
$c$ [$10^{19}~{\rm cm}^{-3}$]	& $-1.54\pm0.34$	& $-1.52\pm0.10$\\ \hline
\end{tabular}
\label{table:fitting}
\end{table}
We now discuss the origin of the 50~meV peak in TaP and the 30~meV peak in TaAs.
The temperature dependence of the absorption intensity can be evaluated as the effective electron number ($N_p$), 
which is obtained by integrating the optical conductivity $\sigma_p(\omega)$ spectra of the peak after subtracting the Drude and low-energy $\omega$-linear components from the original $\sigma(\omega)$ spectra, as plotted in Fig.~\ref{fig:Peak}.
As mentioned above, note that the subtraction of the low-energy $\omega$-linear contribution leaves some ambiguity in the evaluation of these peaks.

As shown in Fig.~\ref{fig:OCnarrow}, the temperature dependence of these peaks seems to be larger than normal interband transitions.
However, the intensity of such low-energy interband transition is strongly suppressed with increasing temperature by thermal screening effect, namely Pauli blocking, which has been observed in the optical spectrum of the typical Dirac semimetal graphite~\cite{Kuzmenko2008}.
Then, the absorption intensity $N_p$ can be described by the following function:
\begin{equation}
N_p(T) = N_{p0}[f_{FD}(-E_F,T)-f_{FD}(E_g-E_F,T)]+c.
\label{eq:Pauliblocking}
\end{equation}
Here, $E_F$ and $E_g$ are the Fermi energy that is the energy difference between the lower Weyl band (LWB) and the chemical potential $\mu$ at $k_x=0$ and the energy difference between the LWB and the upper Weyl band (UWB) connecting a set of Weyl points, respectively (see also inset of Fig.~\ref{fig:Peak}), 
and $N_{p0}$ and $c$ are the peak intensity at $T = 0~{\rm K}$ and temperature-independent background intensity, respectively.
The fitted curves (plotted as dashed and dot-dashed lines in Fig.~\ref{fig:Peak}) represent the obtained peak intensities.
Therefore, the fitting result suggests that the temperature-dependent peak intensities originate from the Pauli blocking effect of the thermal excitation between LWB and UWB.

According to the fitting results, the gap size $E_g$ of $69.1\pm5.4$~meV for TaP and $23.3\pm0.1$~meV for TaAs are similar to the peak energies of each materials, so the fitting might be plausible.
In TaAs, $E_F$ ($11.6\pm0.3$~meV) is almost half of $E_g$.
This suggests that the Weyl points are located very close to $\mu$, if band dispersions of LWB and UWB are symmetrical.
In contrast, in TaP, $E_F$ ($56.8\pm5.0$~meV) is close to $E_g$, which implies that LWB (or UWB) is far away from $\mu$ and UWB (or LWB) is close to $\mu$.

The constant background $c$ of TaP is almost equal to that of TaAs.
This suggests that the ambiguity of the low-energy $\omega$-linear component assumed as the dashed line in Fig.~\ref{fig:OCnarrow} is plausible.

So far, we discuss the far-infrared peak as an absorption owing to the interband transition between LWB and UWB.
However, no peak structure appears for $\hbar\omega\leq100~{\rm meV}$ in the calculated $\sigma(\omega)$ spectra in Fig.~\ref{fig:OCwide}, where only interband transitions are adopted.
This suggests that these peaks may originate not only from interband transitions but also from bulk excitons~\cite{Ueda1986} or from Fermi arc surface states~\cite{Shi2017}.



Note that a similar peak has been observed in another WSM, YbMnBi$_2$~\cite{Chinotti2016}.
In the analysis of YbMnBi$_2$, the corresponding peak was assigned to be due to the interband transition between the LWB and UWB.
Therefore, the peak structure is considered to evidence the existence of a Weyl-type electronic structure, however such large peak has not been expected by theoretical calculations.
To manifest our speculation of the peak origin, further optical investigation of WSMs and theoretical work should be performed.

\subsection{Comparison with band calculations}
\begin{figure}[t]
\begin{center}
\includegraphics[width=0.45\textwidth]{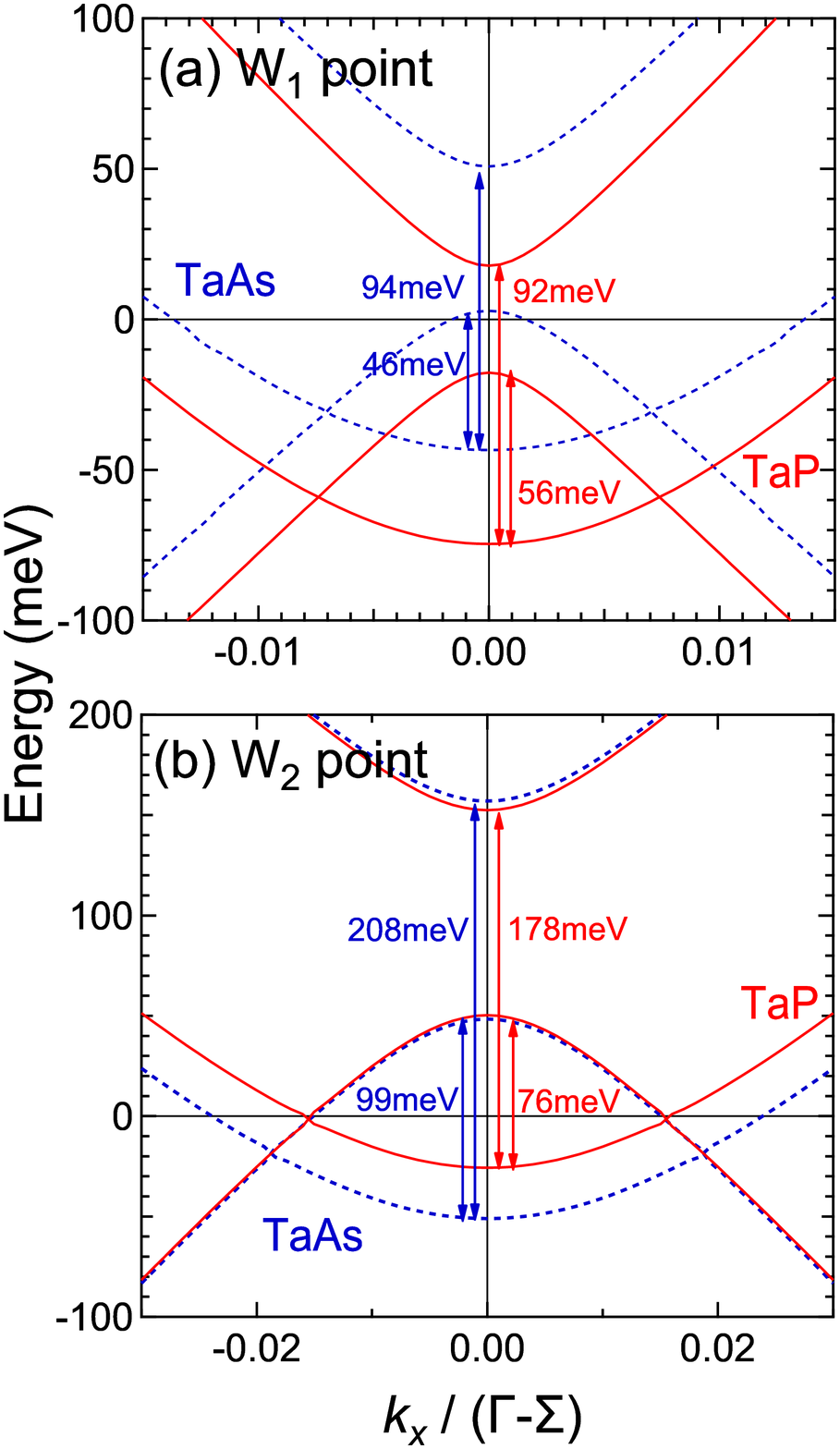}
\end{center}
\caption{
(Color online)
Band dispersions near the $W_1$ (a) and $W_2$ (b) Weyl points of TaP (solid lines) and TaAs (dashed lines) along the $k_x$ axis.
The horizontal axis is normalized by the distance between the $\Gamma$ and $\Sigma$ points.
The zero energy is $E_F$ evaluated from the summation of the electron number.
The center coordinates of the figures for TaP and TaAs are (0, 0.9510$(\Gamma-\Sigma)$, 0) and (0, 0.9413$(\Gamma-\Sigma)$, 0) for $W_1$, and (0, 0.5057$(\Gamma-\Sigma)$, 0.5855$(\Gamma-{\rm Z})$) and (0, 0.5195$(\Gamma-\Sigma)$, 0.5915$(\Gamma-{\rm Z})$) for $W_2$.
}
\label{fig:Weyl}
\end{figure}
Finally, the obtained $\sigma(\omega)$ spectra are compared with the band calculations.
The above discussed bending point of the $\omega$-linear behavior corresponds to the energy difference between the saddle points of UWB and LWB~\cite{Tabert+Carbotte2016}.
Hence, the saddle point separation should be at least the same or \textit{smaller} for TaP (bending point not larger than 85~meV) than for TaAs (bending point at 85~meV).
However, from analyzing the spectral weight of the peaks, the saddle point separation is \textit{larger} for TaP ($69.1\pm5.4$~meV) than for TaAs ($23.3\pm0.1$~meV), 
see $E_g$ in Table~\ref{table:fitting} and inset of Fig.~\ref{fig:Peak}.
Therefore, the two different methods for evaluating the energy separation between the UWB and LWB yield energy sizes for TaP and TaAs in opposite order.
This implies that two kinds of interband transitions between the UWB and LWB states exist.
%
%

The band dispersions near two Weyl points, $W_1$ and $W_2$, are shown in Fig.~\ref{fig:Weyl}.
The band calculations suggest that the band shapes and energies of the UWB and LWB differ for each kind of Weyl points.
Hence, the spectra given by the interband transition between the UWB and LWB are different for $W_1$ and $W_2$, i.e., 
the energy differences between the UWB and LWB for $W_1$ and $W_2$ are 56~meV (46~meV) and 76~meV (99~meV), respectively, in TaP (TaAs).
The size of the band gap near $W_1$ is larger in TaP than in TaAs, which corresponds to the behavior of the bending point.
In contrast, the opposite behavior occurs near $W_2$, which corresponds to the gap size evaluated by the peak.
Therefore, the $\sigma(\omega)$ spectra can be effectively explained by the calculated electronic structure.
However, $E_F$ must change by about $-20~{\rm meV}$ in TaP, considering the appearance of the interband transition near the $W_1$ point in Fig.~\ref{fig:Weyl}(a).
After the correction of $E_F$, the energy of the $W_1$ point becomes about $-40~{\rm meV}$, which is consistent with the value evaluated by the quantum oscillations measurements~\cite{Arnold2016}.
For the case where $E_F$ is shifted down to $-20~{\rm meV}$, the carrier density $N_0$ and the effective carrier mass $m_D/m_0$ are shown in Fig.~\ref{fig:Drude}.
After the change, the crossing temperature of $N_0$ between TaAs and TaP becomes similar to that of $N_D$, and the change in $m_D/m_0$ becomes smaller than before (0.56 at 300~K to 0.3 at 9~K), which is almost consistent with the change in $\Gamma$.
Therefore, these results imply that the actual $E_F$ might be located at $-20~{\rm meV}$.
This is consistent with the band structure of TaP evaluated in NQR experiments~\cite{Yasuoka2016}.

The next higher optical transition that will occur near the $W_1$ point is expected to appear above 90~meV in both materials.
A slight convex shape at $\hbar\omega\sim$~150~meV in the $\sigma(\omega)$ spectrum of TaP and small peaks at $\hbar\omega\sim$~150~meV for TaAs (see Fig.~\ref{fig:OCnarrow}) might originate from the interband transition.

\section{Conclusion}

We have measured the optical conductivity spectra of typical WSMs (Ta$Pn$, $Pn=$~P, As) and compared them with band calculations.
The obtained spectra can be effectively explained by the electronic band structure calculations without electron correlations.
However, a strong temperature-dependent peak can be observed at $\hbar\omega\leq50~{\rm meV}$ in both materials, and this cannot be reproduced by the band calculations.
The temperature dependence of the peak can be explained by the thermal screening of the optical absorption between the saddle points connecting a pair of Weyl points.
However, since the spectral shape cannot be explained by band calculations, an excitonic state of the saddle points or Fermi arc surface states might be adopted.
Owing to the temperature-dependent Drude weight, it has been revealed that the Weyl points of TaP are located far from the Fermi level, whereas those of TaAs are close to $E_F$.
These observations, and the overall agreement with the band calculations, provide important insights into the electronic structure of typical Weyl states in Ta$Pn$.

\section*{Acknowledgments}
We would like to thank Y. Ohtsubo, J. Watanabe, Y. Kayanuma, and J. Carbotte for their fruitful discussion and Y. Nakajima for his support during the optical experiments.
JS acknowledges the support by the International Collaboration Program of Osaka University.
Part of this work was supported by the Use-of-UVSOR Facility Program (BL7B, 2015) of the Institute for Molecular Science.
This work was also supported by a JSPS Grant-in-Aid for Scientific Research (B) (Grant No. JP15H03676).


%
\end{document}